\documentclass[11pt,a4paper]{article}   
\usepackage{amsmath}
\usepackage{amssymb}
\usepackage{mathrsfs}

\usepackage{rotating}

\usepackage[round]{natbib}

\newcommand{\ie}{i.e.\ }

\usepackage[normalem]{ulem}
\usepackage{setspace}
\usepackage{xcolor}
\usepackage{marginnote}

\usepackage{xparse}

\NewDocumentCommand{\marcomm}{mo}
 {
  \IfValueTF{#2}
  {\marginnote[\framebox{\parbox{40pt}{\setstretch{1.0}#1}}]{\framebox{\parbox{40pt}{\setstretch{1.0}#1}}}[#2] }
  {\marginnote[\framebox{\parbox{40pt}{\setstretch{1.0}#1}}]{\framebox{\parbox{40pt}{\setstretch{1.0}#1}}}[0pt] }
 }
 
 \NewDocumentCommand{\revmarcomm}{mo}
 {
  \IfValueTF{#2}
  {\reversemarginpar\marginpar[\color{blue}\framebox{\parbox{33pt}{\setstretch{1.0}\scriptsize\sffamily#1}}]{\vspace{#2}\color{blue}\framebox{\parbox{33pt}{\setstretch{1.0}\scriptsize\sffamily#1}}} }
  {\reversemarginpar\marginpar[\color{blue}\framebox{\parbox{33pt}{\setstretch{1.0}\scriptsize\sffamily#1}}]{\vspace{-5pt}\color{blue}\framebox{\parbox{33pt}{\setstretch{1.0}\scriptsize\sffamily#1}}} }
 }
 
\begin{document}
\title{Hydroelastic interactions between water waves and floating freshwater ice}
\author{A.~Dolatshah$^{1,2}$, 
F.~Nelli$^{2}$,
L.~G.~Bennetts$^{3}$, 
M.~H..~Meylan$^{4}$, 
A.~Alberello$^{3}$,  
J.~P.~Monty$^{5}$, 
A.~Toffoli$^{2}$
\\
{\footnotesize
$^{1}$Faculty of Science, Engineering and Technology, Swinburne University of Technology, Melbourne, VIC 3122, Australia.}
\\
{\footnotesize
$^{2}$Department of Infrastructure Engineering, University of Melbourne, Parkville, VIC 3010, Australia.}
\\
{\footnotesize
$^{3}$School of Mathematical Sciences, University of Adelaide, Adelaide, SA 5005, Australia.}
\\
{\footnotesize
$^{4}$School of Mathematical and Physical Sciences, University of Newcastle, Callaghan, NSW 2308, Australia.}
\\
{\footnotesize
$^{5}$Department of Mechanical Engineering, University of Melbourne, Parkville, VIC 3010, Australia.}
}
\date{\today}
\maketitle

\begin{abstract}
Hydroelastic interactions between regular water waves and floating freshwater ice are investigated using laboratory experiments for a range of incident wave periods and steepnesses.
It is shown that only incident waves with sufficiently long period and large steepness 
break up the ice cover, and that the extent of breakup increases with increasing period and steepness.
Further, it is shown that an increasing proportion of the incident wave propagates through the ice-covered water as period and steepness increase, 
indicating the existence of a positive feedback loop between ice breakup and increased wave propagation.      
\end{abstract}


Hydroelasticity involves deformations of an elastic body in response to hydrodynamic excitations and the reciprocal modification of the excitations due to the body motions. 
It finds application in many fields, including modelling vestibular systems\citep[][]{rabbitt1992hydroelastic}, 
vitreous cortexes\citep[][]{repetto2004simple}, 
ultra-large ships\citep[][]{chen2006review}, 
floating airports\citep[][]{lamas2015review},
slamming problems\citep[][]{Shams2017},
hydrofoil design\citep[][]{Chae2016}
and channel flow\citep[][]{Amaouche2016}.
Most relevant to this study, hydroelasticity is used as standard in modelling interactions between ocean waves  and sea ice covers\citep[][]{williams2013wavea,williams2013waveb}. 

Ocean waves penetrate deep into the ice-covered oceans and break up the ice cover into ice floes with lengths comparable to the wavelengths\cite{squire1995ocean}.
Observations of wave-induced breakup are serendipitous due to the difficulties in making measurements in the harsh and dynamic fringes of the ice-covered ocean, but are becoming more frequent as their potential importance is now better appreciated, particularly due to the changes being experienced by the Earth's sea ice.

Concomitantly, the sea ice attenuates waves over distance, so that the waves only retain the strength to cause breakup up to 100\,kms into the ice cover\citep{asplin2012fracture,kohout2016situ}. 
Field measurements show that long-period waves propagate farther into the ice-covered ocean than short-period waves, and empirical\citep{meylan2014situ} and model\citep{bennetts2010three} propagation--period relationships have been derived.
Less is known about the relationship between wave propagation and the properties of the ice cover, but, notably, \citet{collins2015situ} report a breakup event in the Arctic, and a subsequent increase in wave energy propagation through the broken ice cover. 
 
Contemporary models of wave--ice interactions\citep[][]{williams2013wavea,williams2013waveb} couple a breaking criterion, 
in which strains imposed by waves are measured against a failure strain, with a model of wave energy propagation through the ice-covered ocean that incorporates an attenuation model\citep{bennetts2012calculation,bennetts2012model}. Experimental data is needed to validate and advance coupled breakup/propagation models, in terms of their qualitative and quantitative predictions, and laboratory experiments are likely to play an important role due to the lack of field data, particularly the lack of simultaneous wave and breakup measurements. In this regard, \citet{herman2018floe} analysed the floe size distribution of an ice cover broken by regular waves in a large (72\,m long, 10\,m wide) wave--ice tank, finding that floe sizes cover up to 5 orders of magnitude and have a large-floe cut-off that can be attributed to wave-induced strains. 
However, tests were conducted for very mild wave steepnesses only, and without measuring wave propagation along the ice-cover.

\begin{figure}[h]
\begin{centering}
\setlength{\unitlength}{0.012500in}
\includegraphics[width=0.7\linewidth]{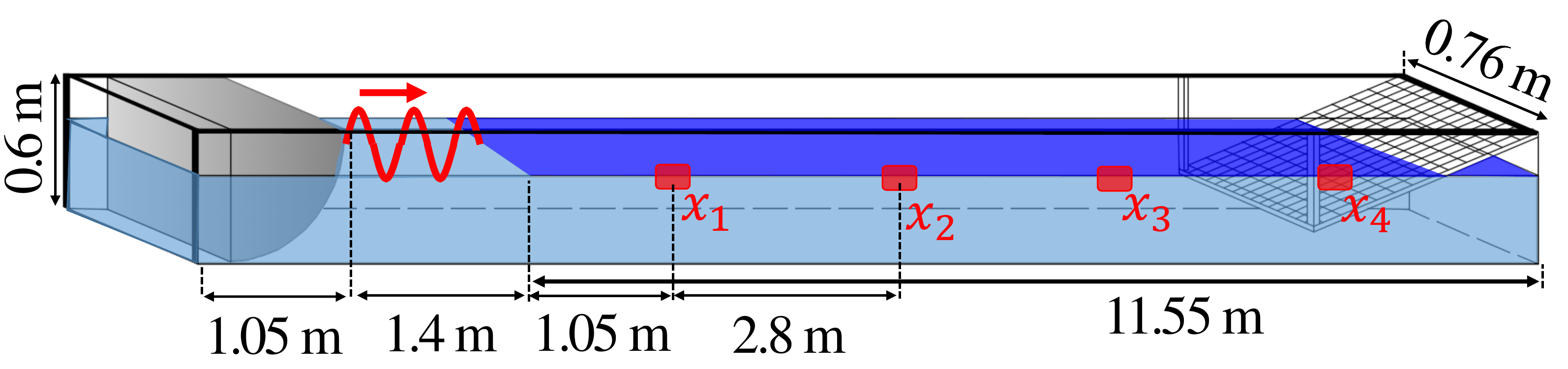}
\caption{Schematic of wave flume, with the wave maker at the left-hand end and the 
beach at the right-hand end. The light-blue is the water and the dark blue is the initial ice cover. Red rectangles indicate the camera locations.}
\label{iceschematic} 
\end{centering}
\end{figure}

In this letter, laboratory experiments on interactions between water waves and freshwater ice are reported, 
as a step towards understanding the complex, coupled processes of wave propagation through ice-covered water and wave-induced breakup of the ice cover. The experiments were conducted at the University of Melbourne in a facility consisting of a wave flume, as shown in Fig.~\ref{iceschematic}, housed inside a refrigerated chamber, where air temperatures can be reduced to $-15^\circ\,\text{C}$. The flume is made out of glass  supported by a wooden frame, ensuring optical access and that the structure experiences minimal contraction or expansion during freezing and defrosting. The flume is 14\,m long, 0.76\,m wide, and was filled with fresh water 0.45\,m deep. It is bounded at one end by a computer-controlled cylindrical wave-maker; and at  the opposite end by a linear beach with slope 1:6, which absorbs incoming wave energy (95\% energy-effective for waves tested). 

\begin{figure}[h]
 \centering
\includegraphics[width=1.0\textwidth]{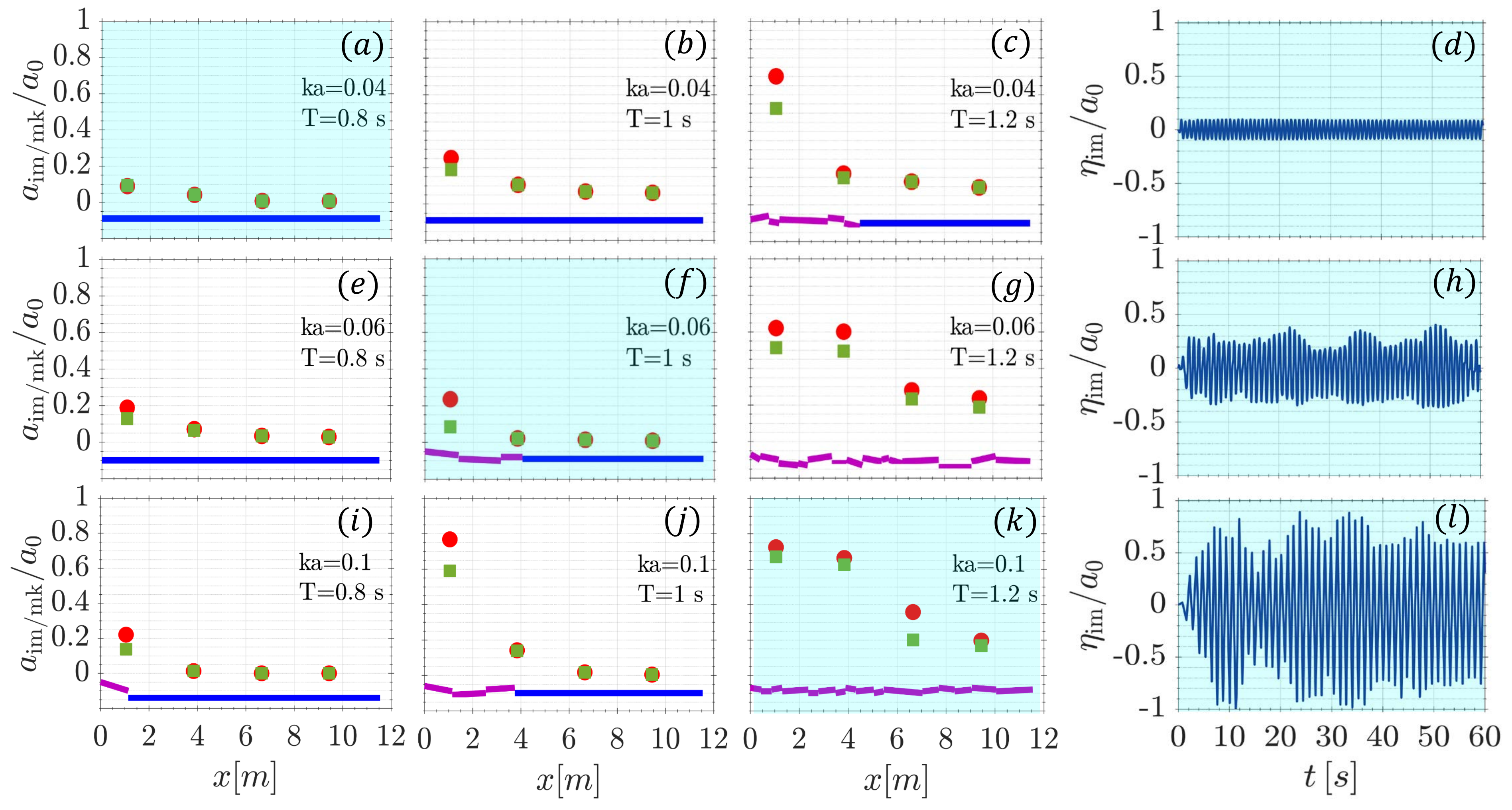}
\caption{(a--c,e--f,i--k)~Mean ice (green squares) and wave (red circles) amplitudes at  $x=x_{i}$ ($i=1$,\dots,4) -- see supplementary material for definitions of parameters. The continuous blue and jagged purple lines at the bottom of panels (a--c,e--f,i--k) indicate the unbroken and broken status of the ice cover respectively at end of each tests; the lengths of broken ice floes is the measured  dimension, while vertical displacements are arbitrary. (d,h,l)~Example water surface elevation time series for the highlighted (light blue background) cases at $x_{1}=1.05$\,m; if no overwash, water and ice surface elevations coincide.}
\label{PlotAttenuationPlusFloeLength} 
\end{figure}

Regular incident wave fields were generated with different periods, $T$, and amplitudes, $a$. Specifically, three wave periods $T=0.8$, 1 and 1.2\,s were used, with corresponding wavelengths $\lambda=1$, 1.56 and 2.1\,m, respectively. For each wave period, three wave amplitudes were chosen to impose wave steepnesses $ka =0.04$, 0.06, and 0.1, where $k=2\pi/\lambda$ is the wavenumber. 

For each incident wave configuration, a floating ice cover was grown, beginning with the water temperature being reduced to $0^\circ\,\text{C}$ by keeping the air temperature at $-1^\circ\,\text{C}$ for 24\,h. The freezing process was then initiated by dropping the air temperature to $-12^\circ\,\text{C}$ for 5\,h, at the end of which the ice cover was $\approx 0.01$\,m thick and covered the water surface along the full flume length. Insulation panels were applied at the sidewalls (from the outside) to prevent ice forming on them.  
Nevertheless, the ice cover at the water surface close to the sidewalls formed a lattice binding it to the walls.
In order to detach the ice from the sidewalls without compromising its mechanical properties, the air temperature was raised to $-1^\circ \text{C}$  for 12\,h before performing each test. 
The air temperature was subsequently maintained at $-1^\circ\,\text{C}$ (with water temperature at $0^\circ\,\text{C}$) to prevent ice melting during the experiments. 
To allow wave generation in open water, ice in the initial 1.4\,m of the flume was removed every 0.25\,h. Within such a short time frame, only a thin, fragile layer of ice formed on the water surface, which could be easily broken up and removed without compromising the mechanical properties of the main cover. 

With the ice at the desired initial condition, the wave maker was used to generate the specified wave field for 60\,s (including a 5\,s run-up, not analysed). In all tests, the incident waves force a layer of water $\approx 3$--$50$\,mm deep onto the surface of the ice at the leading ice edge (see supplementary material), 
similarly noted in previous laboratory tests on wave interactions with plastic floes\cite{mcgovern2014experimental,nelli2017reflection,sree2017experimental} and associated models\cite{skene2015modelling,skene2018water},
and referred to as overwash.
Four cameras with sampling rate of 60\,Hz and resolution of $1280\times{}720$ pixels 
were deployed at distances $x=x_{i}$ $(i=1,2,3,4)$ from the leading ice edge, where $x_{1}=1.05$\,m, $x_{2}=3.85$\,m, $x_{3}=6.65$\,m and $x_{4}=9.45$\,m (see Fig.~\ref{iceschematic}). 
An image processing technique was used to monitor the oscillatory vertical displacements of the surface in contact with air, 
\ie displacements of the air--ice interface is monitored when no overwash is present, 
and displacements of the air--water interface is monitored when overwash is present
(definitions of the surface displacements can be found in the supplementary material).
Video frames were converted from RGB into a grey scale and the water and ice were identified based on their luminance,
and observations are accurate to 1~pixel $\approx 0.02$\,mm. 
Ice displacements were also recorded (with and without overwash) by tracking red markers embedded in the ice during the freezing process and in the same field of view of the cameras.
Accuracy of air--water interface recognition was tested against standard capacitance wave gauges during benchmark runs without the ice cover. Benchmark observations were also used to verify the regular shape of incident wave field, including within the initial 1.4\,m of the flume. 
At the end of all tests, the length of broken floes and the remaining continuous ice cover were recorded using a tape measure. At completion, the ice was melted at air temperature of $3^\circ \text{C}$, before commencing the next test.

Bulk elastic properties of fresh water ice were obtained with the aid of three-point flexural tests. Five samples of ice, with length 0.4\,m, were collected in the flume after wave-induced breakup. Post processing of mechanical tests indicated the Young's modulus of  1--7\,GPa with a 95\% confidence interval, noting that the characteristic length is the basin--field scale for the Young's modulus \cite{Tim80}. 

Fig.~\ref{PlotAttenuationPlusFloeLength} shows the 
amplitudes corresponding to the displacements obtained from the image processing and markers, denoted $a_\textnormal{\small{}im}$ and $a_\textnormal{\small{}mk}$, respectively, 
at each of the observation points along the flume and for each of the nine tests, normalised with respect to the incident wave amplitude, $a_{0}$, as measured in the absence of ice. 
It also shows the ice configuration (broken/unbroken) at the end of each test, and example time series of the normalised water surface displacement from the image processing, $\eta_\textnormal{\small{}im}/a_{0}$, at $x_{1}=1.05$\,m, 
for the tests indicated with light-blue backgrounds in the corresponding row. The original footage of the surface elevation is presented in the supplementary material.

For the least steep and shortest period incident wave tested, $ka=0.04$ and $T=0.8$\,s in Fig~\ref{PlotAttenuationPlusFloeLength}a, the ice surface undulates with waves and the overwash is shallow, meaning the amplitudes $a_\textnormal{\small{}im}$ and $a_\textnormal{\small{}mk}$ are identical (to two decimal places). The amplitudes are reduced significantly at the first observation point compared with the incident wave amplitude, with $a_\textnormal{\small{}im}/a_{0}\approx{}a_\textnormal{\small{}mk}/a_{0}\approx0.09$. Fig.~\ref{PlotAttenuationPlusFloeLength}d shows that, despite the reduction in amplitude, the wave propagating along the ice cover maintains its regular form. It then reduces steadily along the tank, with $a_\textnormal{\small{}im}/a_{0}\approx{}a_\textnormal{\small{}mk}/a_{0}\approx0.01$ at the last observation point, and without breaking the ice.

For the larger steepness, $ka=0.06$ in Fig~\ref{PlotAttenuationPlusFloeLength}e, 
the overwash is deeper, and the  amplitude $a_{\textnormal{\small{}im}}$ is slightly greater than $a_{\textnormal{\small{}mk}}$ 
at the first observation point. The reduction at the first observation point is less than for the smallest steepness, but nonetheless significant, with 
$a_{\textnormal{\small{}im}}/a_{0}\approx0.19$,
which is $\approx{}1.5$ times greater than 
$a_{\textnormal{\small{}mk}}/a_{0}$.
Moreover, the reduction along the ice cover is rapid, so that, at the last observation point, $a_{\textnormal{\small{}im}}/a_{0}\approx0.03$.
The waves do not break the ice but do produce a crack approximately half way along the ice cover.

For the largest steepness, $ka=0.1$ in Fig~\ref{PlotAttenuationPlusFloeLength}i, the ice breaks close to its leading edge after 5\,s, creating a $1.15$\,m-long floe, \ie slightly grater than the incident wavelength. The wave amplitude at the first observation point, in the vicinity of the wave-induced breakup, is larger than for the smaller steepnesses, with 
$a_{\textnormal{\small{}im}}/a_{0}\approx0.22$,
which is $\approx{}1.6$ times greater than $a_{\textnormal{\small{}mk}}/a_{0}$, due to the presence of overwash.
Note that the large bending moments induced by the wave motion cause break up, while the increase of overwash depth is a consequence of break up rather than a cause. The amplitudes measured at the subsequent observation points, in the continuous ice cover, are significantly reduced, with $a_{\textnormal{\small{}im}}/a_{0}\approx0.01$, $0.00$, and $0.00$, respectively. 

For the intermediate wave period, $T=1$\,s, the two steepest incident waves, $ka=0.06$ in Fig~\ref{PlotAttenuationPlusFloeLength}f and 0.1 in Fig~\ref{PlotAttenuationPlusFloeLength}j, break the ice into multiple floes at the leading edge, over a distance slightly greater than two incident wavelengths. The breakup for $ka=0.06$ is relatively slow in comparison to $ka=0.1$, for which the breakup occurs in $<10$\,s, \ie 10 wave periods. For all steepnesses, the amplitudes are generally greater than in the corresponding tests with the shortest period, particularly at the first observation point where the increase is up to a factor of 3.5 (for the largest steepness). Moreover, the oscillations of the water surface becomes less regular, as shown in Fig~\ref{PlotAttenuationPlusFloeLength}h for the $ka=0.06$ case, due to the presence of $\approx 30$\,mm deep overwash (comparable to the incident amplitude $a\approx{}25$\,mm). Amplitudes $a_\textnormal{\small{}im}$ are $1.3$, $2.8$ and $1.3$ times larger than the $a_\textnormal{\small{}mk}$ for $ka=0.04$, $0.06$, and $0.1$, respectively. For all steepnesses, by the third observation point $a_\textnormal{\small{}im}/a_{0}<0.10$, 
and by the fourth observation point $a_\textnormal{\small{}im}/a_{0}\approx 0.06$.

For the longest incident period, $T=1.2$\,s, the extent of wave-induced breakup significantly increases. 
For even the smallest steepness, $ka=0.04$ in Fig~\ref{PlotAttenuationPlusFloeLength}c, ice is broken for $\approx{}2/5$ of the ice cover, \ie slightly greater than two incident wavelengths. 
The first floe broke away from the edge after $<5$\,s $\approx{}4\,T$, but the next breaking event did not occur until $\approx{}30$\,s $=25\,T$. Between the first two breaking events, the overwash reaches the second observation point, so that the overwash amplitude is slightly larger there. 

The two steepest waves, $ka=0.06$ and $0.1$ in Fig.~\ref{PlotAttenuationPlusFloeLength}g and Fig.~\ref{PlotAttenuationPlusFloeLength}k, respectively, break up the entire ice cover, and the breakup reached the far end of the ice cover after only $5$--10\,s ($\approx$4--9 wave periods), noting that further breakup occurred following this.
The wave amplitudes at the second, third and fourth observation points are substantially greater than the smallest steepness case,
with the amplitudes at the second observation point similar to those at the first observation point, and non-negligible amplitudes, $a_\textnormal{\small{}im}/a_{0}>a_\textnormal{\small{}mk}/a_{0}>0.15$, at the fourth observation point. Moreover, the overwash was intense, getting deeper as the tests progressed, generally $\approx{}30$\,mm but up to 50\,mm, and reaching the fourth observation point after $\approx{}40$\,s. 
The overwash depth is up to 5 times the ice thickness, and 
$\approx{}1.5$ times the incident amplitude, $a\approx{}35$\,mm.
The intense overwash created a highly irregular surface elevation, 
as shown in Fig.~\ref{PlotAttenuationPlusFloeLength}l for the steepest incident waves. 

\begin{table}
\begin{center}
\begin{tabular}{ cccccccc  }
   $ka$ & $T$ [s] & $\lambda$ [m] & $a$ [m] & $n_{\text{fl}}$ & $l_{\text{av}}$ [m] & $l_{\text{mn}}$ [m] & $l_{\text{mx}}$ [m]  \\[2pt] \hline
    0.06 & 1.0 & 1.56 & 0.0149 & 3 & 1.40 & 1.05 & 1.75  \\
    0.1 & 1.0  & 1.56 & 0.0248 & 3 & 1.25 & 1.15 & 1.40  \\
    0.04 & 1.2  & 2.1 & 0.0143 & 5 & 0.90 & 0.43 & 2.00 \\
    0.06 & 1.2  & 2.1 & 0.0215 & 14 & 0.83 & 0.40 & 1.50  \\
    0.1 & 1.2  & 2.1 & 0.0358 & 16 & 0.72 & 0.40 & 1.60 \\ 
  \end{tabular}
\caption{Incident wave and break up statistics at the end of the tests in which multiple floes were created: incident wave steepness ($ka$), period ($T$), length ($\lambda$), and amplitude ($a$); number of broken floes ($n_{\text{fl}}$); average floe length ($l_{\text{av}}$); minimum floe length ($l_{\text{mn}}$); and maximum floe length ($l_{\text{mx}}$).}
\label{tab1}
\end{center}
\end{table}

Tab.~\ref{tab1} provides information on the incident wave fields, and ice break up statistics at the end of tests in which multiple floes were created.  
Break up is quantified by the number of broken floes, $n_{\text{fl}}$, and the average (mean), minimum and maximum floe lengths, $l_{\text{av}}$, $l_{\text{mn}}$ and $l_{\text{mx}}$, respectively. As Fig.~\ref{PlotAttenuationPlusFloeLength} indicates,  
the number of broken floes significantly increases from 3--5 in the two intermediate period tests and smallest steepness long-period test, to 14--16 in the two steepest long-period tests.
The mean floe length decreases at a rate of 2.75--3.75\,m per unit steepness, and 2.65--2.85\,m per second period.
The latter contradicts the standard assumption that floe sizes are proportional to wavelengths, as wavelength increases with increasing period. The minimum floe lengths are reasonably consistent with respect to period, 
and, notably, they are identical to one decimal place for both periods. There is a pronounced decrease in the minimum floe lengths of $\approx{}0.7$\,m for increasing period. No trends in maximum floe length are evident, with both increases and decreases as the steepness and period vary. 

\begin{figure}
 \centering
\includegraphics[width=0.5\textwidth]{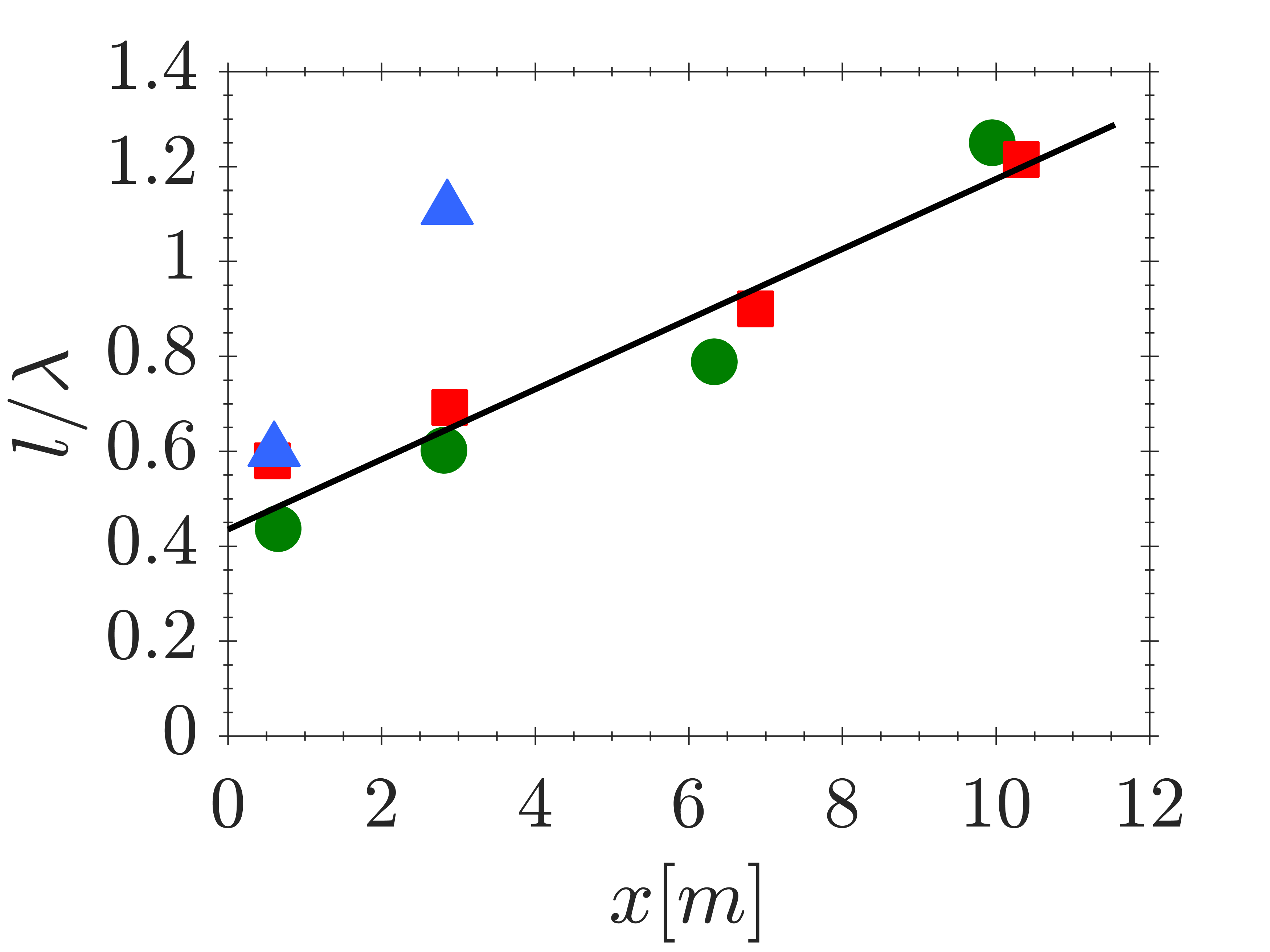}
\caption{Floe size distribution over distance
$T=1.2$\,s with $ka=0.04$ (triangle), $ka=0.06$ (square), $ka=0.1$ (circle). A linear regression for $ka=0.06$--0.1 is overlaid ($-$).}
\label{floesize}
\end{figure}

Fig.~\ref{floesize} shows broken floe lengths, $l$, with respect to distance along the ice cover for the longest-period tests, $T=1.2$\,s, and non-dimensionalised by the incident wavelength $\lambda=2.1$\,m.
The floe size data points are averages over sub-intervals of the ice cover, where the $x$-locations are the interval mid-points. In all tests, the floe lengths increase with distance into the ice cover. For the least steep incident wave, which breaks only 5 floes over $\sim{}2/5$ of the ice cover, the floe lengths increase relatively rapidly from $l/\lambda\approx{}0.6$ close to the ice edge to  $l/\lambda\approx{}1.1$ around $x=3$\,m. The rapid increase is due to one 2\,m-long floe, which is 2.6--4.3 times longer than the surrounding floes. For the two steeper incident waves, which break up the full length of ice cover, the floe sizes close to the leading edge are comparable to those for the least steep case, particularly for $ka=0.06$. However, the floe lengths increase more gently with distance along the ice cover, and there the floe lengths are remarkably consistent for the two different steepnesses. 
The linear regression for the two steeper waves 
\begin{equation}
l/\lambda = 0.0739x + 0.4347,
\end{equation}
is overlaid, and
indicates that floe lengths increase at a rate $\approx{}0.16$\,m per metre into the ice cover.

In summary, results were reported from laboratory experiments on interactions between regular incident water waves and freshwater ice. It was shown that short-period, small-steepness incident waves travel only a short distance into the ice-covered water without breaking the ice. As the incident waves get longer and steeper, the waves propagate farther into the ice cover and rapidly break up the ice cover over an increasing distance at least 5 wavelengths for the longest waves tested, 
which would be on the order of kilometres or more at field scale,
and over only a few minutes.
Most striking, a sharp transition was noted to rapid breakup of the entire ice cover and wave propagation along the full length of ice-covered water, indicating  the existence of a positive feedback loop between increased breakup and increased propagation. 
Moreover, the increased propagation and breakup led to extensive overwash,
submerging some of the broken floes, 
which would likely affect floe melt rates in the field.
The results cast new light on hydroelastic wave--ice interactions, 
showing, for the first time, simultaneous observations of wave propagation and wave-induced ice breakup, and paving the way for less idealised investigations into the natural, field-scale phenomenon. 


\section*{Supplementary material}
See supplementary material for the complete electronic laboratory experiment.


\section*{Acknowledgements}

The Australian Research Council funded the facility (LE140100079). 
AD was support by a Swinburne University  Postgraduate Research Award and Nortek AS. MM was supported by EPSRC grant no. EP/K032208/1 and was also partially supported by a grant from the Simons Foundation. AA and AT were funded by the Antarctic Circumnavigation Expedition Foundation and Ferring Pharmaceuticals. AD, FN, AA and AT acknowledge support from the Air-Sea-Ice Lab Project.

\end{document}